\documentclass[epj,english,twocolumn]{webofc}
\usepackage[varg]{txfonts}   % Web of Conferences font
\usepackage{hyperref}        % hyperref package
\usepackage{graphicx}
\usepackage{array,longtable,calc}
\usepackage{threeparttable}

\newcommand{\comment}[1]{}

\woctitle{Heavy Ion Accelerator Symposium 2019}

\begin{document}

\title{Emerging nuclear collectivity in $^{\bf{124-130}}$Te}

\author{B.~J.~Coombes\inst{1} \and
        A.~E.~Stuchbery\inst{1} \and
        J.~M.~Allmond\inst{2} \and
        A.~Gargano\inst{3} \and
        J.~T.~H.~Dowie\inst{1} \and
        G.~Georgiev\inst{4} \and
        M.~S.~M.~Gerathy\inst{1} \and
        T.~J.~Gray\inst{1} \and
        T.~Kib\'edi\inst{1} \and
        G.~J.~Lane\inst{1} \and
        B.~P.~McCormick\inst{1} \and
        A.~J.~Mitchell\inst{1} \and
        N.~J.~Spinks\inst{1} \and
        B.~P.~E.~Tee\inst{1}
}

\institute{Department of Nuclear Physics, Research School of Physics, The Australian National University, Canberra, ACT, 2601, Australia
\and
           Physics Division, Oak Ridge National Laboratory, Oak Ridge, Tennessee 37831, USA
\and
           Istituto Nazionale di Fisica Nucleare, Complesso Universitario di Monte S. Angelo, Via Cintia, I-80126 Napoli, Italy
\and
           CSNSM, CNRS/IN2P3; Universit\'e Paris-Sud, UMR8609, F-91405 ORSAY-Campus, France
          }

\abstract{%
The emergence of nuclear collectivity near doubly-magic $^{132}$Sn was explored along the stable, even-even $^{124-130}$Te isotopes. Preliminary measurements of the $B(E2;4^{+}_{1}\rightarrow2^{+}_{1})$ transition strengths are reported from Coulomb excitation experiments primarily aimed at measuring the $g$~factors of the $4^{+}_{1}$ states. Isotopically enriched Te targets were excited by 198-205~MeV $^{58}$Ni beams. A comparison of transition strengths obtained is made to large-scale shell-model calculations with successes and limitations discussed.
}
\maketitle
\section{Introduction}\label{Intro-Sec}
Studies of the emergence of collective excitations across isotopic chains give essential information on the degrees of freedom important in creating nuclear collectivity and the nature of the collectivity that develops. The stable even isotopes from $^{120}$Te to $^{130}$Te are two protons away from the $Z$=50 closed shell and the highest mass isotopes are close to the $N$=82 closed neutron shell. They can be compared to the Cd isotopes ($Z$=48) with two proton holes, which have been more extensively studied. There is an ongoing discussion on the nature of the collectivity that, in the Cd isotopes, is traditionally associated with anharmonic vibrations~\cite{Chamoli2011,Stuchbery2016,Coombes2019,Perez2016,Garrett2010,Garrett2019}.

The Te isotopes show increasing collectivity as they depart further from doubly-magic $^{132}$Sn. The large number of stable Te isotopes allows an extensive and systematic study of the emergence of collectivity across a single isotopic chain by Coulomb excitation.

Recent work on the Xe isotopes ($Z$=54) has suggested that the $g$~factors and $E2$ transition strengths for states above the first $2^{+}$ states converge more slowly to the collective limits than the $2^{+}_{1}$ states and have suggested that collectivity begins with the $2^{+}_{1}$ state before moving to the higher excited states~\cite{Peters2019}. A natural next step in the study of such phenomena is the Coulomb excitation of the Te isotopes, which are closer to the $Z$=50 shell closure. The Te isotopes lie between the Xe and Sn isotopes and have been investigated previously~\cite{Vanhoy2004,Hicks2008,Hicks2012,Hicks2017,Sabri2016,Stone2005,Stuchbery2007,Stuchbery2007b} with comparison made to the harmonic vibrational model, however, the transition strengths between the $2^{+}_{1}$ and $4^{+}_{1}$ have not previously been reported in $^{128,130}$Te. The previous measurement in $^{126}$Te has a large ($\sim$40\%) uncertainty. In the unstable isotopes $^{132,134}$Te, the $6^{+}$ states are isomeric and have previously measured lifetimes and $g$~factors; also the $4^{+}_{1}$ state has a known lifetime in $^{134}$Te~\cite{Goodin2008,Wolf1976}. These properties are largely consistent with a $(\pi g_{7/2})^2$ seniority structure~\cite{Allmond2011,Stuchbery2013}. The systematic development of collectivity is observable in $B(E2;2^{+}_{1}\rightarrow0^{+}_{1})$ values, which show a gradual increase in transition strength towards the neutron mid-shell. In this work, we aim to determine the feasibility of using data collected in a recent transient-field $g$-factor measurement~\cite{Coombes3000} to investigate the changing collectivity of the $4^{+}_{1}\rightarrow2^{+}_{1}$ transition strengths in the stable, even-even $^{124-130}$Te isotopes including the previously unreported values in $^{128,130}$Te.

\section{Experiment}\label{Exp-Sec}
Enriched $^{124-130}$Te targets of $\sim$0.6~mg/cm$^2$ with $\sim$5~mg/cm$^2$ iron and $\sim$6-9~mg/cm$^2$ copper backing layers were bombarded with 198-205~MeV $^{58}$Ni ions at a beam current of $\sim$1.5~pnA. Backscattered beam particle-$\gamma$ coincidences were measured with an XIA Pixie-16 digital pulse processor~\cite{XIA}. The experiment was performed with beams from the ANU 14UD Pelletron accelerator and with the ANU hyperfine spectrometer~\cite{Stuchbery2020}. The target was kept at a constant temperature of $\sim$4~K by a Sumitomo RDK-408D cryocooler to help prevent beam-induced damage to the target. Cooling the target was necessary due to the relatively low melting point of Te (449.5~$^{\circ}$C). The $\gamma$ rays were measured with four HPGe clover detectors, each with four segments, from the CLARION array~\cite{Gross2000} placed 11.3~cm from the target position in the horizontal plane. Addback of coincident $\gamma$ rays in different segments of each clover detector was performed. Each $\gamma$-ray detector had FWHM of $\sim$2~keV at 1~MeV. Particles were detected in two silicon photodiode detectors with widths of 25.17~mm and heights of 9.25~mm, placed 16.2~mm upstream from the target position and 4.6~mm vertically above and below the beam axis. Outputs from the particle detectors were first processed because they showed large amplitude and low-frequency oscillations under beam which could be readily filtered out by the analog electronics modules. (We have yet to fully understand the origin of this behaviour.) The processed signals were input to the digital data acquisition system. Details of beam energies, target thickness and $\gamma$-ray detector angles are given in Table~\ref{TeDatasets}. Detector angles were chosen to be close to the angle of maximum sensitivity in the $g$-factor measurement.
\renewcommand{\arraystretch}{1.2}
\begin{table}[t!]
  \caption{Experimental details: $E_{B}$ is the $^{58}$Ni beam energy and L$_{\rm{Te}}$ is the target thickness. The angles $\theta_{\gamma}$ are the polar angles to the centres of the Clover detectors. }
\begin{tabular}{ c c c c c c }
\hline
Isotope &  Run    &   $E_{B}$    &   L$_{\rm{Te}}$    & \multicolumn{2}{c}{$\theta_{\gamma}$}   \\
        &         &    (MeV)    &    (mg/cm$^2$)   & Front & Back \\ \hline

$^{124}$Te & A & 200    & 0.42 & $\pm65^{\circ}$ & $\pm125^{\circ}$ \\
$^{126}$Te & B & 205    & 0.59 & $\pm65^{\circ}$ & $\pm125^{\circ}$ \\
$^{128}$Te & C & 205    & 0.57 & $\pm65^{\circ}$ & $\pm125^{\circ}$ \\
$^{128}$Te & D & 205    & 0.57 & $\pm65^{\circ}$ & $\pm115^{\circ}$ \\
$^{130}$Te & E & 198    & 0.72 & $\pm65^{\circ}$ & $\pm125^{\circ}$ \\
$^{130}$Te & F & 205    & 0.72 & $\pm65^{\circ}$ & $\pm115^{\circ}$ \\ \hline
\end{tabular}
\label{TeDatasets}
\end{table}
\renewcommand{\arraystretch}{1.0}

\section{Results and Discussion}\label{Res-Sec}
\begin{figure}[!tb]
\centering
\includegraphics[width=\columnwidth]{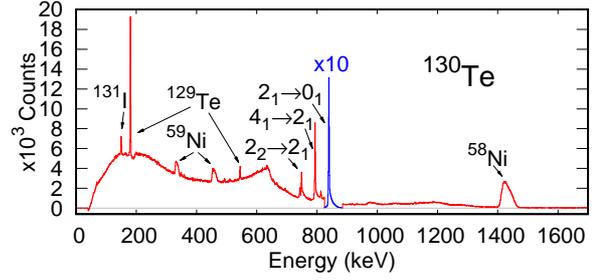}
\caption{Energy spectrum taken at $65^{\circ}$ in Run F as defined in Table~\ref{TeDatasets} (enriched $^{130}$Te target). The decay of the first excited state has been scaled down by a factor of 10.}
\label{ExampleSpectrumFull}
\end{figure}

\begin{figure}[!tb]
\centering
\includegraphics[width=\columnwidth]{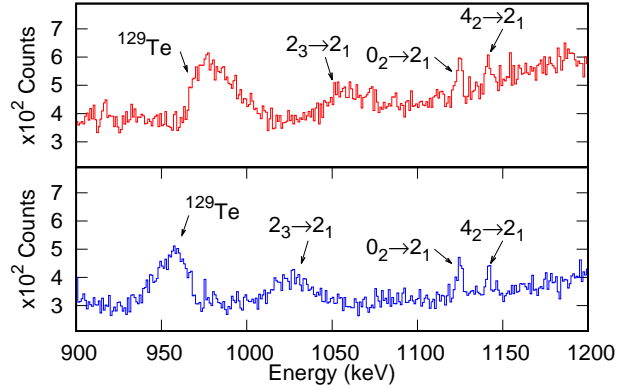}
\caption{Spectra for the region from 900 to 1200~keV in Run F as defined in Table~\ref{TeDatasets} (enriched $^{130}$Te target). The top and bottom panels display data from $65^{\circ}$ and $115^{\circ}$ detectors, respectively.}
\label{ExampleSpectrum1000}
\end{figure}

The transitions observed in the Coulomb excitation measurements are listed in Tables~\ref{Te124transtable}-\ref{Te130transtable}. Several transitions were significantly broadened by in-flight Doppler shift, causing overlap with other transitions. As some of the transitions are not well separated, the intensities, especially for the weaker transitions, can be difficult to extract. Figures~\ref{ExampleSpectrumFull} and \ref{ExampleSpectrum1000} show examples of the measured energy spectra.

Experimental data were analysed using the semi-classical Coulomb-excitation code \textsc{Gosia}~\cite{Czosnyka1983}. The stopping powers of Ziegler~\cite{Ziegler1985} were used where required. Transition strengths were extracted relative to the known $2^{+}_{1}\rightarrow0^{+}_{1}$ transition strengths~\cite{Katakura2008,Katakura2002,Elekes2015,Singh2001}. Analysis was also performed to measure transition strengths relative to the recently measured $^{58}$Ni transition strength~\cite{Allmond2014}. The resulting $2^{+}_{1}\rightarrow0^{+}_{1}$ transition strengths agree with the previously measured values within $\sim$10\%, with the exception of $^{124}$Te, which is found to be 20\% lower. The experimental beam energies were chosen to maximize excitation for the simultaneous $g$-factor measurement. Coulomb excitation is termed safe when there is no significant overlap of the projectile and target wavefunctions. A common method to ensure this is to maintain a nuclear surface separation of 5~fm~\cite{Czosnyka1983}. In the present experiment, for a head-on collision, this corresponds to $\sim$69\% of the Coulomb barrier with the Coulomb barrier as defined in Ref.~\cite{Samuel1968} and a nuclear radius parameter of $r_{0}=1.25$~fm. The excitation occurs at $\sim$72-75\% of the barrier and is therefore not purely safe Coulomb excitation, which increases the uncertainty in the measured $B(E2)$ values as the nuclear effects can interfere constructively or destructively. The magnitude of the (typically destructive) nuclear interference can be estimated by a \textsc{Ptolemy}~\cite{MacFarlane1978} distorted wave Born approximation calculation. A worst-case-scenario calculation suggested that up to a 40\% difference could be caused by the nuclear interactions at the scattering angles used in the present work. It is not clear that this effect would cancel in the relative transition-strength analysis. It is possible to perform an analysis using coupled calculations including the nuclear effects~\cite{Samuel1968}, however these calculations have not been performed.

Although the data taken in the present experiment were insufficient to determine the signs of the matrix elements involved in Coulomb excitation, the results can be sensitive to the relative signs. This effect can be large and cannot be determined without measurements that include multiple scattering angles or beam energies.

Despite these difficulties, an exploratory analysis of the excited-state Coulomb excitation was performed. The aim of this analysis was to determine the $4^{+}_{1}\rightarrow2^{+}_{1}$ matrix elements, while other matrix elements were allowed to vary where a measurement of the relevant yield was possible. There are a number of previously measured lifetimes of higher excited states and mixing ratios of the transitions between higher excited states~\cite{Vanhoy2004,Hicks2008,Hicks2012,Katakura2008,Katakura2002,Elekes2015,Singh2001} which were used to constrain matrix elements that could not be determined in the current experiment. The signs of all matrix elements, including those determined from previously measured lifetimes, were taken to be those predicted by shell-model calculations, as were the matrix elements involving weakly populated states.

Shell-model calculations were performed with the large-scale shell-model code \textsc{Antoine}~\cite{Caurier2005}. Calculations were performed with a $^{100}$Sn core and with two body matrix elements from the CD-Bonn potential. Empirical effective charges of $e_p=1.7e$ and $e_n=0.9e$ were used in the calculations, which are similar to those used in other studies in the region~\cite{Peters2019,Teruya2015}. Both protons and neutrons were allowed to occupy the full ($g_{7/2},d_{5/2},d_{3/2},s_{1/2},h_{11/2}$) model space. The transition strengths predicted by the shell-model calculations and experimentally determined $B(E2)$ values are shown in Table~\ref{BE2resulttable}. The signs predicted by shell-model calculations for the $E2$ matrix element between the $2^{+}_{1}$ and $4^{+}_{1}$ states denoted $M_{23}$ (relative to a positive $2^{+}_{1}\rightarrow0^{+}_{1}$ matrix element) are given next to the shell-model values in Table~\ref{BE2resulttable}. The effect of changing the sign of $M_{23}$ can be large and is given in the same table for context. This sensitivity is due to the significant excitation strength from paths through higher excited states, such as through excitations to the $4^{+}_{2}$ state. The shell-model calculations are reasonably successful in reproducing the transition strengths for the $2^{+}_{1}\rightarrow0^{+}_{1}$ transitions. However, there is a consistent underestimation of $\sim$10\%. The $4^{+}_{1}\rightarrow2^{+}_{1}$ transition strengths determined in the present work agree well with the shell-model calculations close to $^{132}$Sn, however, the calculations do not capture the increase in transition strength suggested by the present data as the number of neutron-holes increases away from $^{132}$Sn.

\renewcommand{\arraystretch}{1.1}
\begin{table}
  \centering
  \caption{Observed transitions in $^{124}$Te. Level energies, spins and parities taken from Ref.~\cite{Katakura2008}.}
\begin{tabular}{ c c c c c }
  \hline
$I^{\pi}_{i}$ &    E$_{i}$     &  E$_{\gamma}$    & $I^{\pi}_{f}$  & E$_{f}$         \\
            &     (keV)      &      (keV)     &              &      (keV)     \\ \hline
$2^{+}_1$    &     602.7      &      602.8     &  $0^{+}_1$    &        0       \\
$4^{+}_1$    &    1248.6      &      645.7     &  $2^{+}_1$    &     602.7      \\
$2^{+}_2$    &    1325.5      &      722.4     &  $2^{+}_1$    &     602.7      \\
            &                &      1325.7    &  $0^{+}_1$    &        0       \\
$0^{+}_2$    &    1657.3      &      1054.7    &  $2^{+}_1$    &     602.7      \\
$6^{+}_1$    &    1747.0      &       498.3    &  $4^{+}_1$    &    1248.6      \\
$0^{+}_3$    &    1882.9      &       556.9    &  $2^{+}_2$    &    1325.5      \\
$4^{+}_2$    &    1957.9      & $\sim$710      &  $4^{+}_1$    &    1248.6      \\
            &                & $\sim$1350     &  $2^{+}_1$    &     602.7      \\ \hline
\end{tabular}
\label{Te124transtable}
\end{table}

\begin{table}
  \centering
  \caption{Observed transitions in $^{126}$Te. Level energies, spins and parities taken from Ref.~\cite{Katakura2002}.}
\begin{tabular}{ c c c c c }
  \hline
$I^{\pi}_{i}$ &    E$_{i}$     &  E$_{\gamma}$    & $I^{\pi}_{f}$  & E$_{f}$         \\
            &     (keV)      &      (keV)     &              &      (keV)     \\ \hline
$2^{+}_1$    &     666.4      &      666.4     &  $0^{+}_1$    &        0       \\
$4^{+}_1$    &    1361.4      &      694.9     &  $2^{+}_1$    &     666.4      \\
$2^{+}_2$    &    1420.2      &      753.8     &  $2^{+}_1$    &     666.4      \\
            &                &      1420.2    &  $0^{+}_1$    &        0       \\
$6^{+}_1$    &    1776.2      &       414.7    &  $4^{+}_1$    &    1361.4      \\
$0^{+}_2$    &    1873.4      &      1207.3    &  $2^{+}_1$    &     666.4      \\
$2^{ }_3$    &    2045.2      &      2045.2    &  $0^{+}_1$    &        0       \\ \hline
\end{tabular}
\label{Te126transtable}
\end{table}

\begin{table}
  \centering
  \caption{Observed transitions in $^{128}$Te. Level energies, spins and parities taken from Ref.~\cite{Elekes2015}.}
\begin{tabular}{ c c c c c }
  \hline
$I^{\pi}_{i}$ &    E$_{i}$     &  E$_{\gamma}$    & $I^{\pi}_{f}$  & E$_{f}$         \\
            &     (keV)      &      (keV)     &              &      (keV)     \\ \hline
$2^{+}_1$    &     743.2      &      742.9     &  $0^{+}_1$    &        0       \\
$4^{+}_1$    &    1497.0      &      753.6     &  $2^{+}_1$    &     743.2      \\
$2^{+}_2$    &    1520.0      &      776.4     &  $2^{+}_1$    &     743.2      \\
            &                &      1520.3    &  $0^{+}_1$    &        0       \\
$6^{+}_1$    &    1811.1      &       314.0    &  $4^{+}_1$    &    1497.0      \\
$0^{+}_2$    &    1978.8      &      1235.6    &  $2^{+}_1$    &     743.2      \\
$4^{+}_2$    &    2027.8      &       530.7    &  $4^{+}_1$    &    1497.0      \\
$4^{+}_2$    &    2027.8      &      1284.6    &  $2^{+}_1$    &     743.2      \\ \hline
\end{tabular}
\label{Te128transtable}
\end{table}

\begin{table}
  \centering
  \caption{Observed transitions in $^{130}$Te. Level energies, spins and parities taken from Ref.~\cite{Singh2001}.}
\begin{tabular}{ c c c c c }
  \hline
$I^{\pi}_{i}$ &    E$_{i}$     &  E$_{\gamma}$    & $I^{\pi}_{f}$  & E$_{f}$         \\
            &     (keV)      &      (keV)     &              &      (keV)     \\ \hline
$2^{+}_1$    &     839.5      &      839.4     &  $0^{+}_1$    &        0       \\
$2^{+}_2$    &    1588.3      &      748.4     &  $2^{+}_1$    &     839.5      \\
            &                &      1588.0    &  $0^{+}_1$    &        0       \\
$4^{+}_1$    &    1633.0      &      793.3     &  $2^{+}_1$    &     839.5      \\
$2^{+}_3$    &    1885.7      & $\sim$1045     &  $2^{+}_1$    &     839.5      \\
$(0)^{+}_2$  &    1964.8      &      1125.2    &  $2^{+}_1$    &     839.5      \\
$4^{+}_2$    &    1981.5      &      348.6     &  $2^{+}_1$    &    1633.0      \\ \hline
\end{tabular}
\label{Te130transtable}
\end{table}
\renewcommand{\arraystretch}{1.0}

\renewcommand{\arraystretch}{1.2}
\begin{table*}[h!]
  \centering
  \begin{threeparttable}
    \centering
    \caption{Transition strengths in $^{124-130}$Te. Data from Nuclear Data Sheets~\cite{Katakura2008,Katakura2002,Elekes2015,Singh2001} and the present work. Transition strengths measured in this work are given without uncertainties. A 20\% uncertainty is assumed for all measured transition strengths. Experimental values are given next to the signs of the matrix elements predicted by the shell-calculations (SM). Details of the shell-model calculations are given in the text.}
    \label{BE2resulttable}
\begin{tabular}{ c c c c c c c c c }
  \hline
Isotope   & Sign($M_{23}$) & \multicolumn{2}{c}{$B(E2; 2^{+}_{1}\rightarrow0^{+}_{1})$} & \multicolumn{3}{c}{$B(E2; 4^{+}_{1}\rightarrow2^{+}_{1})$} & \multicolumn{2}{c}{$B(E2; 6^{+}_{1}\rightarrow4^{+}_{1})$} \\
          &               &             \multicolumn{2}{c}{(W.u.)}                   & \multicolumn{3}{c}{(W.u.)}                  &           \multicolumn{2}{c}{(W.u.)}             \\
          &               &Exp                 &         SM        &    Present    & Previous                   &          SM        &     Exp       &     SM      \\ \hline

$^{124}$Te &      $+$      &                    &                   & 48            &                            &                    &               &                      \\
$^{124}$Te &      $-$      & 31.1(5)            &      26           & 54            &     35.8(16)~\tnote{a}     &      30            & 27            &     17               \\
$^{126}$Te &      $+$      & 25.4(7)            &      23           & 31            &     34(16)~\tnote{b}       &      25            & 17.8(6)       &     14               \\
$^{126}$Te &      $-$      &                    &                   & 36            &                            &                    &               &                      \\
$^{128}$Te &      $+$      & 19.68(18)          &      18           & 19            &                            &      19            & 9.7(6)        &     10               \\
$^{128}$Te &      $-$      &                    &                   & 26            &                            &                    &               &                      \\
$^{130}$Te &      $+$      &                    &                   & 18            &                            &                    &               &                      \\
$^{130}$Te &      $-$      & 15.1(3)            &      14           & 14            &                            &      13            & 6.1(3)        &      7               \\ \hline
\end{tabular}
\end{threeparttable}
\begin{tablenotes}
\item[a] $^{\rm{a}}$ Ref.~\cite{Saxena2014}
\item[b] $^{\rm{b}}$ Ref.~\cite{Stokstad1967}
\end{tablenotes}
\end{table*}
\renewcommand{\arraystretch}{1.0}

\begin{figure}[!tb]
\centering
\includegraphics[width=\columnwidth]{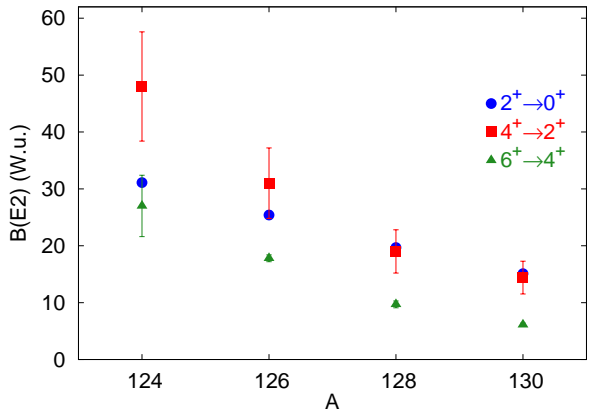}
\caption{Systematics of measured $B(E2)$ values across the Te isotopes. Data from this work and Nuclear Data Sheets~\cite{Katakura2008,Katakura2002,Elekes2015,Singh2001}.}
\label{BE2Systematics}
\end{figure}

Figure~\ref{BE2Systematics} shows the transition strengths between yrast states in the Te isotopes. The measured $4^{+}_{1}\rightarrow2^{+}_{1}$ transition strengths increase at a similar rate to the previously known $2^{+}_{1}\rightarrow0^{+}_{1}$ values, increasing more rapidly further away from the shell closure at $N=82$. The harmonic vibrational limit of $B(E2;4^{+}_{1}\rightarrow2^{+}_{1})/B(E2;2^{+}_{1}\rightarrow0^{+}_{1})=2$ has not yet been reached at $^{124}$Te. The lower mass isotopes $^{120,122,124}$Te have known transition strengths between the $4^{+}_{1}$ and $2^{+}_{1}$ states~\cite{Saxena2014}. The only overlapping measurements between these and present studies are for the $^{124}$Te $B(E2;4^{+}_{1}\rightarrow2^{+}_{1})$ and $B(E2;2^{+}_{2}\rightarrow2^{+}_{1})$ values. The preliminary measured ratio of $B(E2;4^{+}_{1}\rightarrow2^{+}_{1})/B(E2;2^{+}_{1}\rightarrow0^{+}_{1})=1.7(3)$ differs at the $1.8\sigma$ level from the previously reported value ($1.16(5)$). The discrepancy is not so large as to be unreasonable; however, it must be resolved to understand the nature of the developing collectivity. There is a similar difference in the $B(E2;2^{+}_{2}\rightarrow2^{+}_{1})/B(E2;2^{+}_{1}\rightarrow0^{+}_{1})$ values between the same previous experiment and the present work. The present value of 1.9(4) differs from the previously determined value of 1.12(18) by $1.8\sigma$. In the present analysis, the inferred $2^{+}_{2}\rightarrow2^{+}_{1}$ transition strength is particularly sensitive to the transition strengths between weakly excited states. It is worth noting that the present measurement compares well with transitions out of the $6^{+}$ states where previous transition strengths are known between the $6^{+}_{1}$ and $4^{+}_{1}$ states. These values may be compared in Tables~\ref{BE2resulttable}~and~\ref{HigherBE2resulttable}. As stated previously, the Coulomb excitation here is not purely safe, which may explain the differences between the present and previous work.

The measured transition strengths to higher-excited states are summarized in Table~\ref{HigherBE2resulttable}. Other excited states observed were too weakly-excited or not clearly separated to adequately determine $B(E2)$ values.

\renewcommand{\arraystretch}{1.2}
\begin{table}
  \centering
  \caption{Transition strengths in $^{124-130}$Te. A 20\% uncertainty is assumed for all measured transition strengths. Experimental (Exp) and shell-model (SM) values are presented.}
\begin{tabular}{ c c c c c c c }
  \hline
Isotope    & E$_{i}$ & $I^{\pi}_{i}$ & E$_{f}$ & $I^{\pi}_{f}$ & \multicolumn{2}{c}{$B(E2)$}  \\
           & (keV)  &              & (keV) &              & \multicolumn{2}{c}{(W.u.)}   \\
           &        &              &       &              &     Exp       &      SM      \\ \hline
$^{124}$Te  & 1325.5 & $2^{+}_{2}$   & 602.7  & $2^{+}_{1}$   &  58           & 32           \\
$^{124}$Te  & 1747.0 & $6^{+}_{1}$   & 1248.6 & $4^{+}_{1}$   &  25           & 17           \\
$^{126}$Te  & 1420.2 & $2^{+}_{2}$   & 666.4  & $2^{+}_{1}$   &  34           & 28           \\
$^{126}$Te  & 1776.2 & $6^{+}_{1}$   & 1361.4 & $4^{+}_{1}$   &  18           & 14           \\
$^{128}$Te  & 1520.0 & $2^{+}_{2}$   & 743.2  & $2^{+}_{1}$   &  26           & 14           \\
$^{128}$Te  & 1811.1 & $6^{+}_{1}$   & 1497.0 & $4^{+}_{1}$   &  15           & 10           \\
$^{130}$Te  & 1588.3 & $2^{+}_{2}$   & 839.5  & $2^{+}_{1}$   &  12           &5.4           \\
$^{130}$Te  & 1964.8 & $(0)^{+}_{2}$ & 839.5  & $2^{+}_{1}$   & 0.7           &1.4           \\ \hline
\end{tabular}
\label{HigherBE2resulttable}
\end{table}
\renewcommand{\arraystretch}{1.0}

Neutron transfer reactions were observed in runs B-F, clearly showing that direct nuclear effects are present and therefore that the Coulomb excitation cannot be assumed safe; the present results must be considered preliminary. Measurements at safe energies are clearly required.

Additional measurements are anticipated. The present data allow for informed planning of Coulomb excitation measurements on the Te isotopes including the number and identity of excited states expected in such measurements as well as enabling reliable estimates of the beam time required for each isotope.

\section{Conclusion}\label{Conclusion-Sec}
The $4^{+}_{1}\rightarrow2^{+}_{1}$ transition strengths in $^{124-130}$Te have been estimated from Coulomb excitation data obtained in a recent $g$-factor measurement. Despite the beam energies being above those considered `safe', and evidence of nuclear interactions from both calculations and observed transfer reactions, new results for $^{128,130}$Te were obtained that show agreement with shell-model calculations. The $B(E2;4^{+}_{1}\rightarrow2^{+}_{1})/B(E2;2^{+}_{1}\rightarrow0^{+}_{1})$ ratio increases from $^{130}$Te towards $^{124}$Te, as neutrons are removed from the $N=82$ shell-closure. Some discrepancy between the present and previous measurements of transition strengths in $^{124}$Te has been observed, with the shell-model calculations in better agreement with the previous measurement~\cite{Saxena2014}. However, in neither case does $B(E2;4^{+}_{1}\rightarrow2^{+}_{1})/B(E2;2^{+}_{1}\rightarrow0^{+}_{1})$ in $^{124}$Te reach the vibrational limit. The present results obtained as a by-product of $g$-factor measurements show the trends in the onset of collectivity and highlight the importance to perform a comprehensive set of precise and reliable $B(E2)$ measurements in the Te isotopes extending across all of the stable isotopes from $^{120}$Te to $^{130}$Te.

\end{document}